\documentclass[12pt,a4paper,oneside,headings=small]{scrartcl}
\usepackage[utf8]{inputenc}
\usepackage[english]{babel}
\usepackage{graphicx,geometry}
\usepackage{fancyhdr,mathptmx,float}
\usepackage[T1]{fontenc}

\usepackage{helvet,array}
\newcolumntype{L}[1]{>{\raggedright\let\newline\\\arraybackslash\hspace{0pt}}m{#1}}
\usepackage[x11names]{xcolor}
\usepackage{booktabs}
\usepackage{parskip}
\usepackage{upgreek}
\usepackage{subfigure}
\usepackage{epstopdf}
\usepackage{amsmath}
\usepackage[comma,authoryear]{natbib}
\usepackage{tikz}
\usetikzlibrary{shapes, shadows, arrows, positioning,patterns}
\usetikzlibrary{decorations.pathreplacing,calc}

\fancypagestyle{firstPage}{%
\fancyhead[L]{}
\fancyfoot[L]{{\textit{\fontsize{9}{7}\selectfont 13$^{th}$ International Symposium on Hazards, Prevention and Mitigation of Industrial Explosions\\
Braunschweig, GERMANY - July 27-31, 2020}}}
\fancyfoot[R]{\vspace{-3mm}\includegraphics[trim=0cm 0cm 0cm 0cm,clip=true,width=0.09\textwidth]{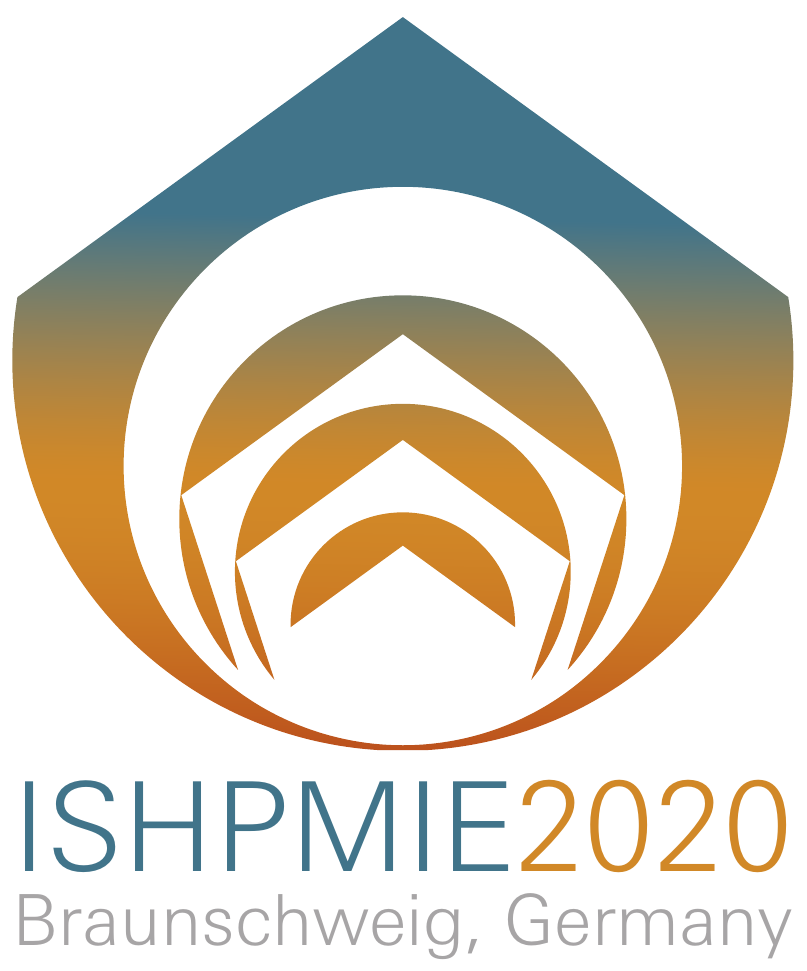}}

}

\setkomafont{section}{\normalfont\bfseries}
\RedeclareSectionCommand[
  beforeskip=-1\baselineskip,
  afterskip=.1\baselineskip]{section}
\setkomafont{subsection}{\normalfont\itshape}
\RedeclareSectionCommand[
  beforeskip=-1\baselineskip,
  afterskip=.1\baselineskip]{section}
\setkomafont{subsubsection}{\normalfont\itshape}
\RedeclareSectionCommand[
  beforeskip=-1\baselineskip,
  afterskip=.1\baselineskip]{section}

\addto\captionsenglish{
        
        }
\usepackage[format=plain,
            labelfont=it,
            textfont=it]{caption}

\begin{document}
\thispagestyle{firstPage}

\begin{center}
\textbf{\Large Measurement of the deposit formation during pneumatic transport of polydisperse PMMA powder}\\[15pt]
Nuki Susanti$^{a,b}$, Holger Grosshans$^{a,b}$\\[5pt]
$^a$ Physikalisch-Technische Bundesanstalt (PTB), Braunschweig, Germany\\
$^b$ Institute of Apparatus- and Environmental Technology, Otto von Guericke University Magdeburg, Germany\\[5pt]
E-mail: {\color{blue} \textit{holger.grosshans@ptb.de}}\\[5pt]
\end{center}

\section*{Abstract}

Deposits which form during the pneumatic transport of particles are a frequent source of explosions in industrial facilities.
Thus, to contribute to the operational safety of industrial plants, we explore the deposit formation of powder flows with our new experimental test-rig. 
To this end, the particle flow was analyzed optically using a square-shaped transparent pipe as a measuring section. 
We characterized the deposit pattern and quantified the fraction of particles that deposit depending on the flow conditions. 
The parameters under consideration included the flow Reynolds number, the powder loading, the duct material, and the ambient conditions (temperature, humidity). 
More specifically, we investigated the transport of polydisperse PMMA particles through pipes made of PMMA and PC.

\medskip
\noindent
\textbf{Keywords:} \textit{pneumatic conveying, deposits, electrostatics, particle-laden flow, experiment}

\section{Introduction}
\label{intro}

When powders are conveyed pneumatically through pipes or ducts, they form deposit layers on component surfaces. 
These deposits are a frequent source of explosions in industrial particle flow facilities~\citep{Eck03}.
For example, if the powder is charged by triboelectric effects, a deposit represents a local accumulation of electrostatic energy, which may lead to hazardous spark discharges.
Also, an external heat source such as welding at the outside of the pipe may initiate the deposit to smolder.
Moreover, it was recently observed that deposited dust in an enclosed pipeline can trigger a reverse explosion \citep{song2019} and the re-entrainment of deposited particles can cause secondary explosions \citep{Eck03}. 
Also, particle deposition along the transporting pipeline causes a reduced cross-sectional area and blockage which might lead to a reduced operation efficiency~\citep{klinzing1981,Kli18}.
Thus, an increased understanding of the physical mechanisms involved in particle deposition will help to design safe and efficient particle transport systems.

In dilute flow pneumatic conveying, which is the focus of the present study, particles are known to adopt preferential positions due to the inhomogeneity of the flow inside the pipe~\citep{Eat94} and electrostatic forces~\citep{Gro18d,Gro18e}. 
The particle concentration becomes even more complicated when the conveying line is of a non-cylindrical shape, such as a square or rectangular duct~\citep{Noo16}. 
Further, pneumatic transport is known to generate high charge due to severe particle-wall collisions~\citep{bailey1984,Gro17a,Gro16a,Gro16b,Gro16f}. 
Those issues make a seemingly simple process highly convoluted since a wide range of factors influences particle motion, dispersion, preferential concentration, and subsequent deposition.

Particle deposition from turbulent flow in non-cylindrical enclosures has been widely studied, both by means of numerical simulation \citep{Wang2019,yao2011,Winkler2006,Gro19e,Gro19b} and experimental investigation \citep{Sippola2004,Kaftori1995,Matsusaka1998}. 
\citet{Wang2019}, as well as \citet{yao2011}, used large eddy simulation (LES) combined with a Lagrangian approach of one-way coupling to simulate deposition at almost the same range of particle sizes (50~$\upmu$m--500~$\upmu$m) and Reynolds numbers (10\,000-250\,000). 
They found that particles have preferential deposition locations depending on size and Reynolds number, either in the center or in the corner of the duct. 
\citet{Sippola2004} experimentally measured small particle deposition (1~$\upmu$m--16~$\upmu$m) at relatively low air velocities (2.2~m/s--8.8~m/s) in a steel ventilation and an insulated duct. 
They pointed out the deposition rate increases due to surface roughness. 
The experimental work of \citet{Kaftori1995} using water as conveying fluid and particles ranging from 100~$\upmu$m--900 $\upmu$m revealed a relation of the particle motion and, thus, deposition with near-wall coherent vortices.

Despite its obvious importance, the detailed mechanism of deposit formation remains unknown such as which forces mainly contribute and where the deposit tends to form. 
This can be attributed to the complex interaction of the fluid mechanics, particle dynamics, and various adhesive forces acting on the particles. 
Consequently, the main objective of the present study is to design, based on our experience with a previous facility \citep{Gro17b} a new experimental test-rig to investigate particle deposition in a square duct. 
The presented experimental facility was designed with the aim to maximize consistency with our complimentary numerical simulations~\citep{Gro18d,Gro19d}.
Then, the influence of the Reynolds number, particle mass flow rate, and relative air humidity (RH) and temperature was investigated.
In this experiment, first particles and a duct of the same material (PMMA) were used to minimize the triboelectric charging effect which might influence deposition. 
Subsequently, in the second part of the study, a measurement is conducted to investigate the influence of particle deposition due to triboelectric charging by using a different duct material (PC).

\section{Experimental set-up}

\begin{figure*}[b]
\centering
\begin{tikzpicture}
\node[anchor=south west,inner sep=0] (A) at (0,0) {\includegraphics[width=.8\textwidth]{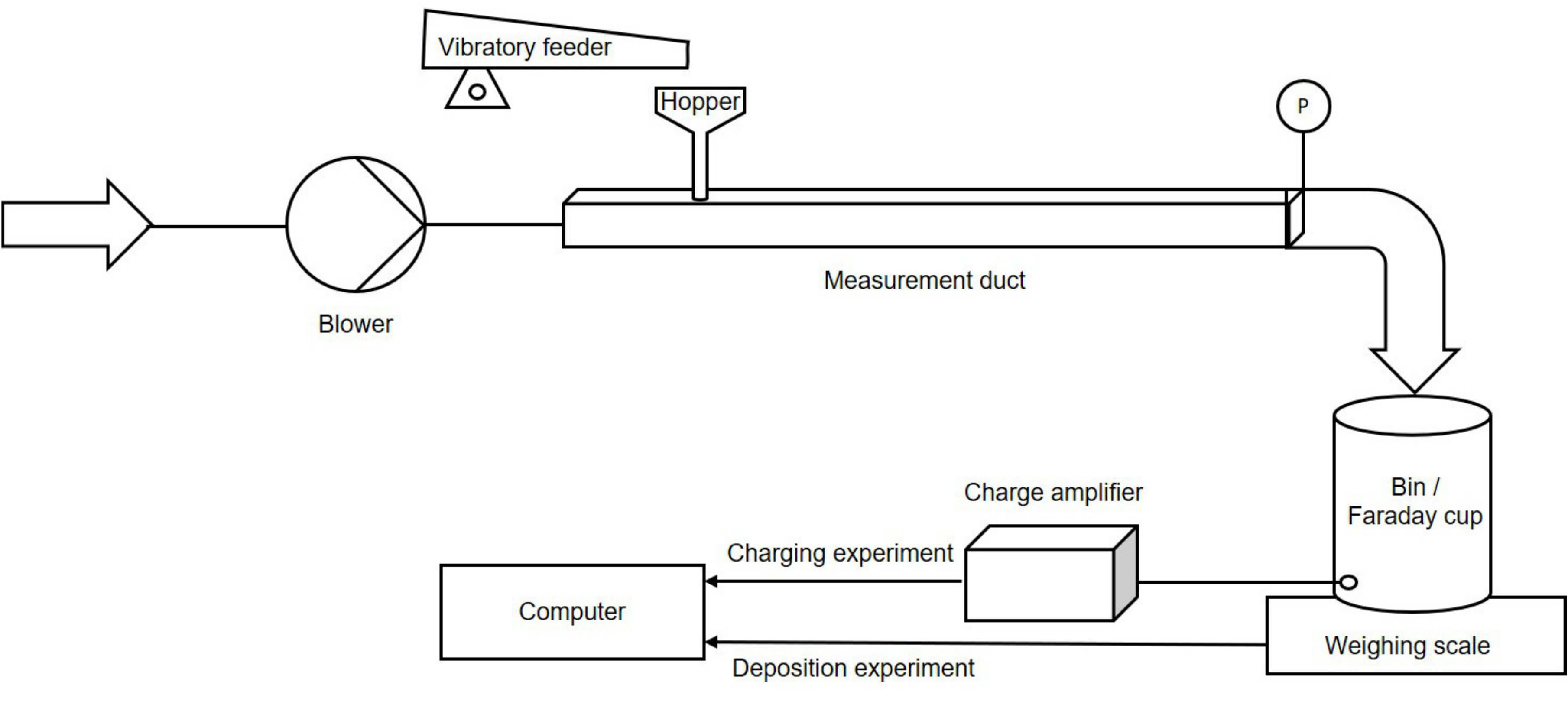}} ;
\draw [<->,thick] (14,5) node[anchor=south,left]{$y$} -- (14,4) -- (15,4) node[anchor=south,below]{$x$} ;
\end{tikzpicture}
\caption{Schematic diagram of the new experimental setup.}
\label{sketch}      
\end{figure*}

The schematics of the new experimental setup  at  Physikalisch-Technische Bundesanstalt (PTB) Braunschweig, Germany, is depicted in Fig.\,\ref{sketch}.
It consists of an air blower (Moro MHR 452) equipped with a frequency converter (Danfoss FC 51) that is connected to a square duct (PMMA ID:45 mm or PC ID: 46 mm) of a length of 1.8~m.
At the end of the duct, a pitot tube anemometer (PCE-HVAC 2) is placed to measure the mean bulk air velocity using the duct traversing technique (Log-Tchebychef). 
Point P in the schemata represents an airflow measuring point. 
Particles are fed into the duct via a vibratory feeder (Retsch DR100) and transported to the bin where the particle mass leaving the duct is weighed and recorded every 60 seconds. 
Thus, the particle mass fraction deposited in the duct can be calculated as
\begin{equation}
\mathrm{deposition \: fraction} = \dfrac{\dot{m}_\mathrm{in} - \dot{m}_\mathrm{out}}{\dot{m}_\mathrm{in}} .
\end{equation}
In the above equations $\dot{m}_\mathrm{in/out}$ are the particle mass flow rate input and output in kg/hr.

\begin{figure*}[t]
\centering
\includegraphics[width=.6\textwidth]{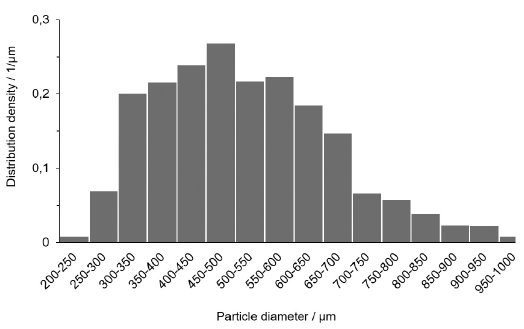}
\caption{Distribution of the volumetric frequency of the PMMA powder XP 200, reprinted from \citet{Gro17b}.}
\label{ParticleDistribution}      
\end{figure*}

We used the same PMMA particles provided by KFG GmbH Biebesheim which were already used in previous work \citep{Gro17b}.
The particle size distribution is plotted in Fig.~\ref{ParticleDistribution}
The experiments were not conducted in the climate chamber, thus, the day-to-day temperature and RH were always carefully considered and recorded.

\section{Results and discussion}

\subsection{Characterization of the airflow}

\begin{figure}[b]
\centering
\subfigure[]{\includegraphics[width=.45\textwidth]{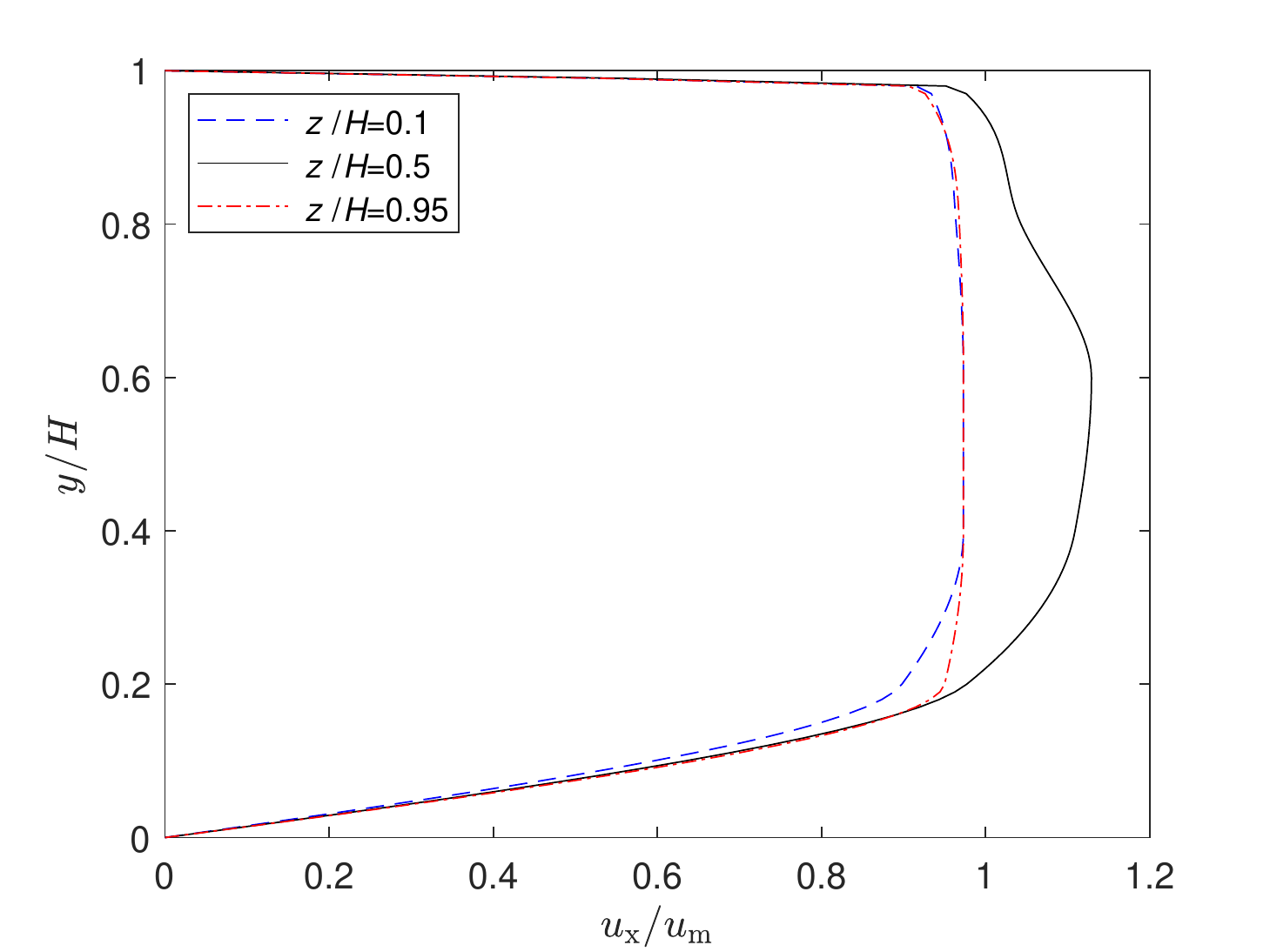}}\qquad
\subfigure[]{\includegraphics[width=.45\textwidth]{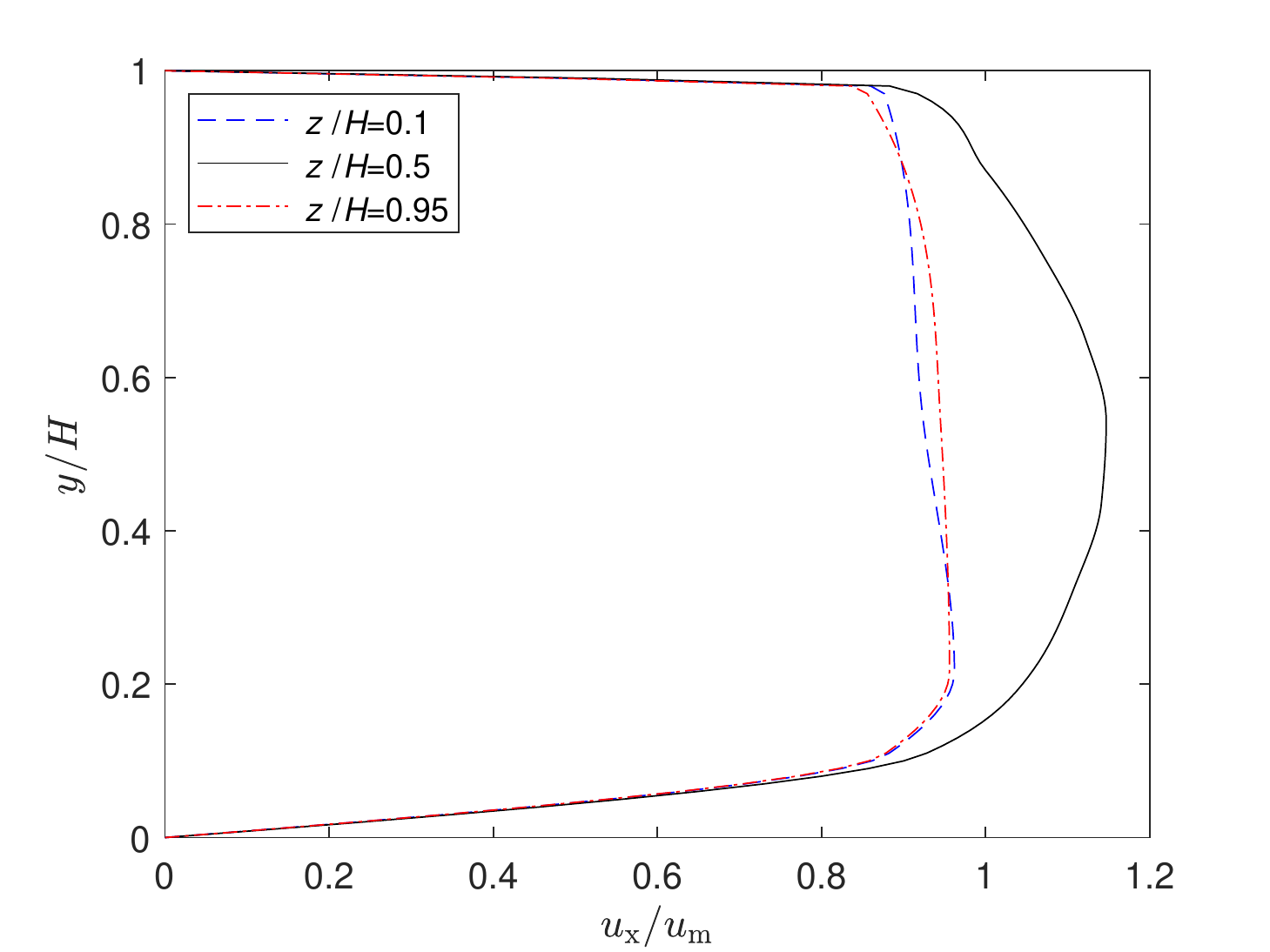}}
\caption{Air velocity profiles at three different cross-sections inside the (a) PMMA duct at $Re=$~12705 and inside the (b) PC duct at $Re=$~25945.}
\label{Velocity Profile}
\end{figure}

Since a pitot tube can only measure the velocity at one particular point, a method to observe the velocity distribution and, thus, the average flow velocity is needed. 
To this end, we used the pitot-static traverse technique \citep{Cheong2001} as proposed by ISO standard 3966.
According to the Log-Tchebychef method the duct is divided into several small squares and the velocity in each corner of the small square is measured.
In order to get a good average velocity, a minimum of 25 measurements is required.
For a duct of a height less than 30~inches (76.2~cm), 5 traversal lines are required.
\cite{Peszynski2018} compared this method to the numerical integration curve resulting in a 2.2\% difference.

The profile of the flow in the PMMA duct is plotted in Fig.\,\ref{Velocity Profile}(a) and in the PC duct in Fig. \ref{Velocity Profile} (b). 
This air velocity measurement is taken at the end of the duct located at 1.8 m from the air inlet point respectively 1.5 m from the particle feeding point (see the experimental sketch) and expressed in terms of the time-averaged velocity.
The measured streamwise air velocity, $u_\mathrm{x}$, and the distance from the wall, $y$ and $z$, are normalized using the mean bulk velocity, $u_\mathrm{m}$, and the inner height of the duct, $H$, respectively. 
Thus, $Re$ is given by
\begin{equation}
   Re = \frac{H \,u_\mathrm{m}}{\nu} \, ,
\end{equation}\\
where \(\nu\) is kinematic viscosity.

\subsection{Influence of the environmental conditions on the deposition velocity}

An important quantity in the following analysis is the deposition velocity which represents the minimum conveying air velocity required to transport particles without settling or forming any stationary or resting deposition layer.
This quantity indicates the operation range of a specific conveying system.
In our experiments, we defined the depoision velocity as the velocity at which particles halt within the first minute of the transport process and form a deposition layer at the bottom of the duct.
We observed that at an air velocity close to this point, the flow appears to be unstable which manifests by the intermittent appearance of particle clusters. 
As the air velocity is further reduced, particles start to settle and form a deposition layer.

\begin{figure}[b]
\subfigure[]{\includegraphics[width=.47\textwidth]{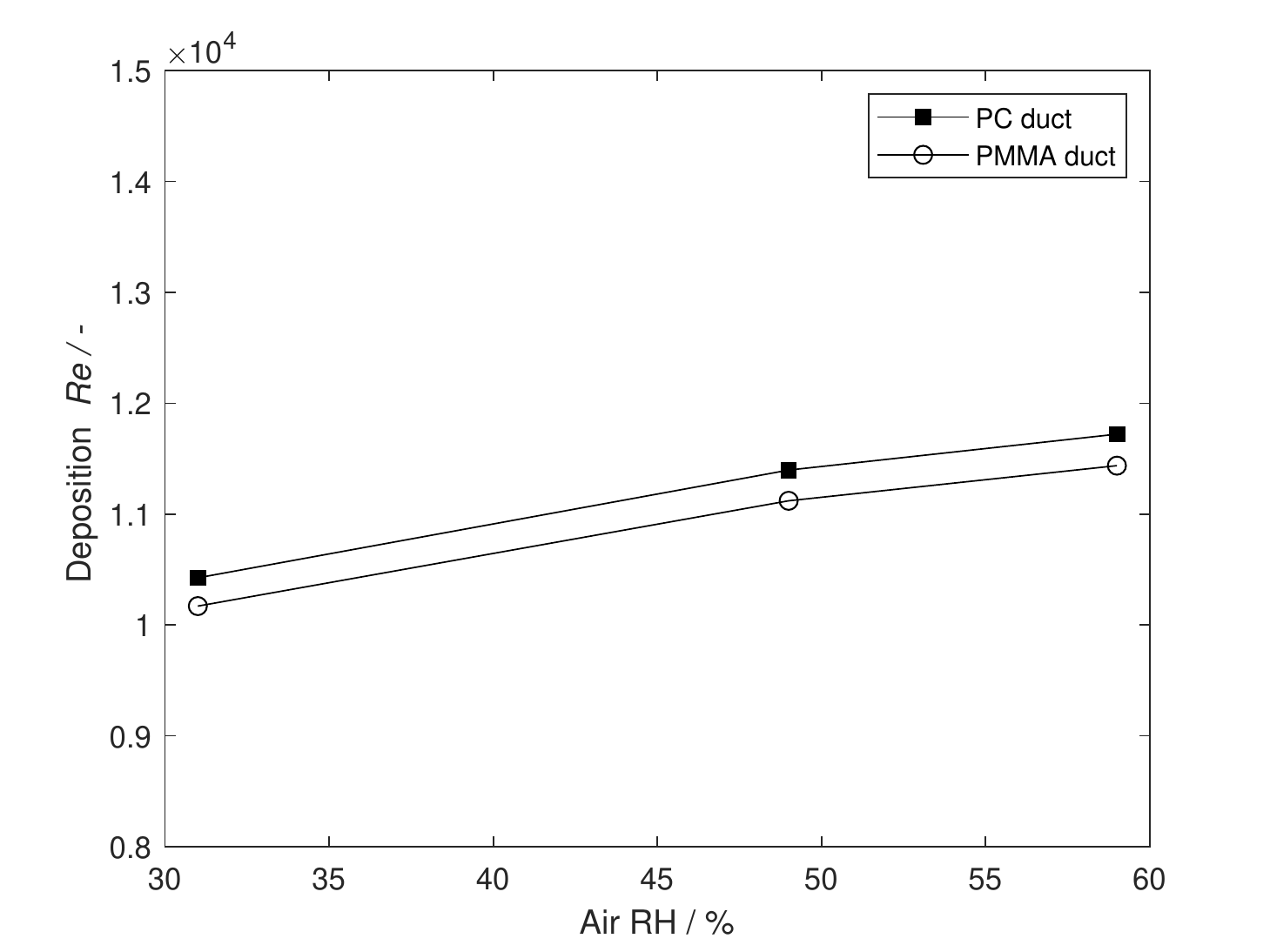}\label{Deposition Velocity Poly}}
\subfigure[]{\includegraphics[width=.47\textwidth]{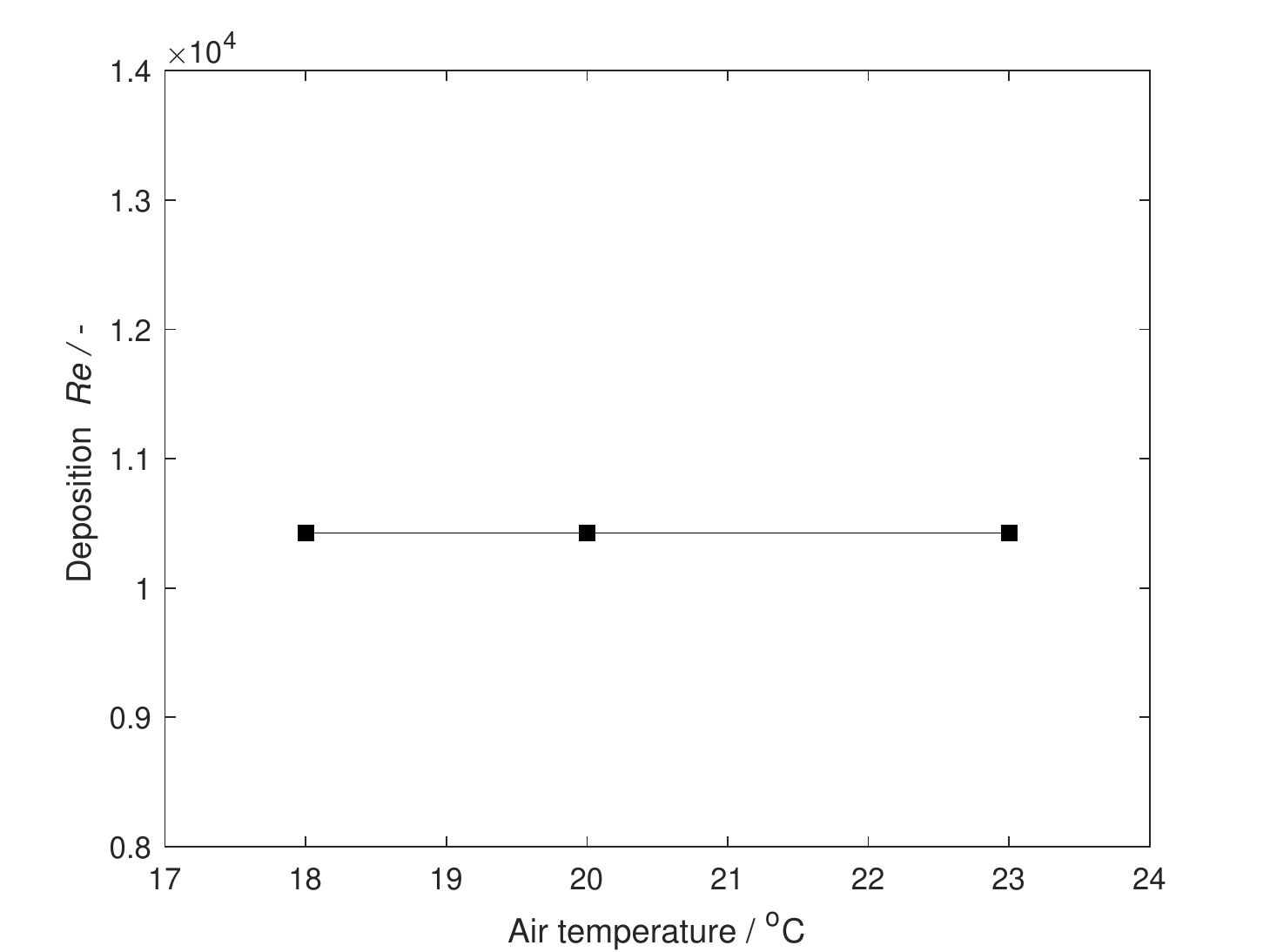}\label{Deposition Velocity Poly Temp}}
\caption{Influence of the air (a) RH and (b) temperature on the deposition velocity of polydisperse particles.}
\label{Deposition Velocity Poly All}
\end{figure}

It is well known that air RH and temperature may affect particle deposition \citep{Han2011,he1998}.
Thus, we first examine the influence of RH and temperature on the deposition velocity and fraction.
Experiments are initially performed to capture the general behavior.
For that purpose, a constant particle flow rate of 3 kg/hr was adjusted.
Three different RHs (31\%, 49\%, and 59\%) were tested while the room temperature was 18~$^{\circ}$C--23~$^{\circ}$C.
The results shown in Fig.\,\ref{Deposition Velocity Poly} indicate that for our coarse polydisperse particles the RH is not of significant influence on the deposition velocity.
Only a slight increase of the deposition velocity occurs as air RH increases drastically causing a comparable deposition fraction.
To examine the temperature effect tests were performed at three temperatures (18$^{\circ}$C, 20$^{\circ}$C, and 23$^{\circ}$C) at fixed RH (31\%) using the PMMA duct.
The constant deposition velocity plotted in Fig.\,\ref{Deposition Velocity Poly Temp} indicates a negligible effect of the temperature at least for relatively large particles.
\citet{he1998} confirmed that the temperature predominantly influences particle transport and deposition in turbulent flow when the particles are rather small, namely less than 0.1 $\upmu$m (1 $\upmu$m according to \citet{Han2011}).
Therefore, the effect of temperature will not be considered for the entire discussion of the present work.

\subsection{Influence of $Re$ and the particle mass flow rate on particle deposition}

\begin{figure*}[b]
\centering
\subfigure[]{\includegraphics[width=.47\textwidth]{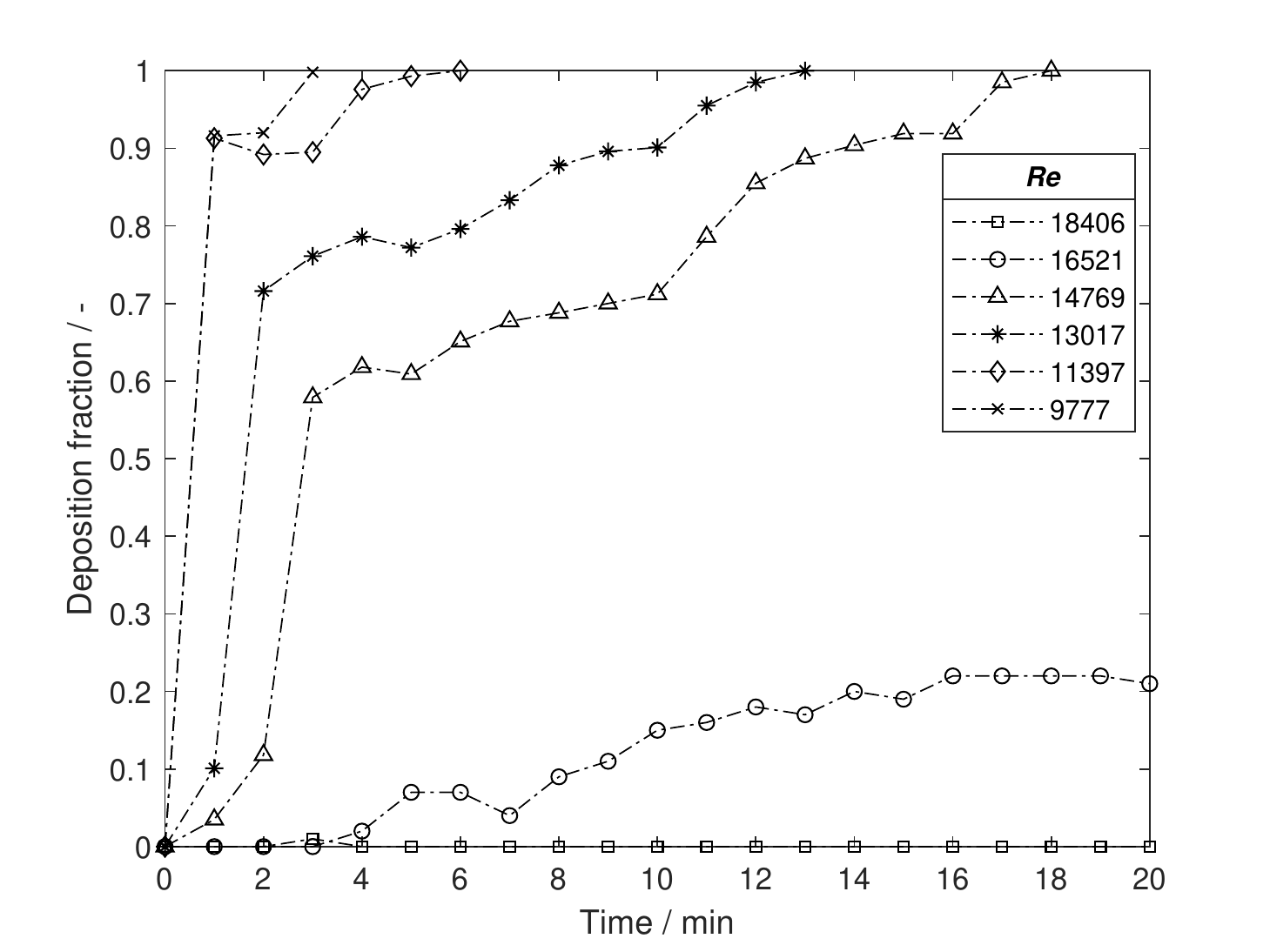}\label{Poly1}}\quad
\subfigure[]{\includegraphics[width=.47\textwidth]{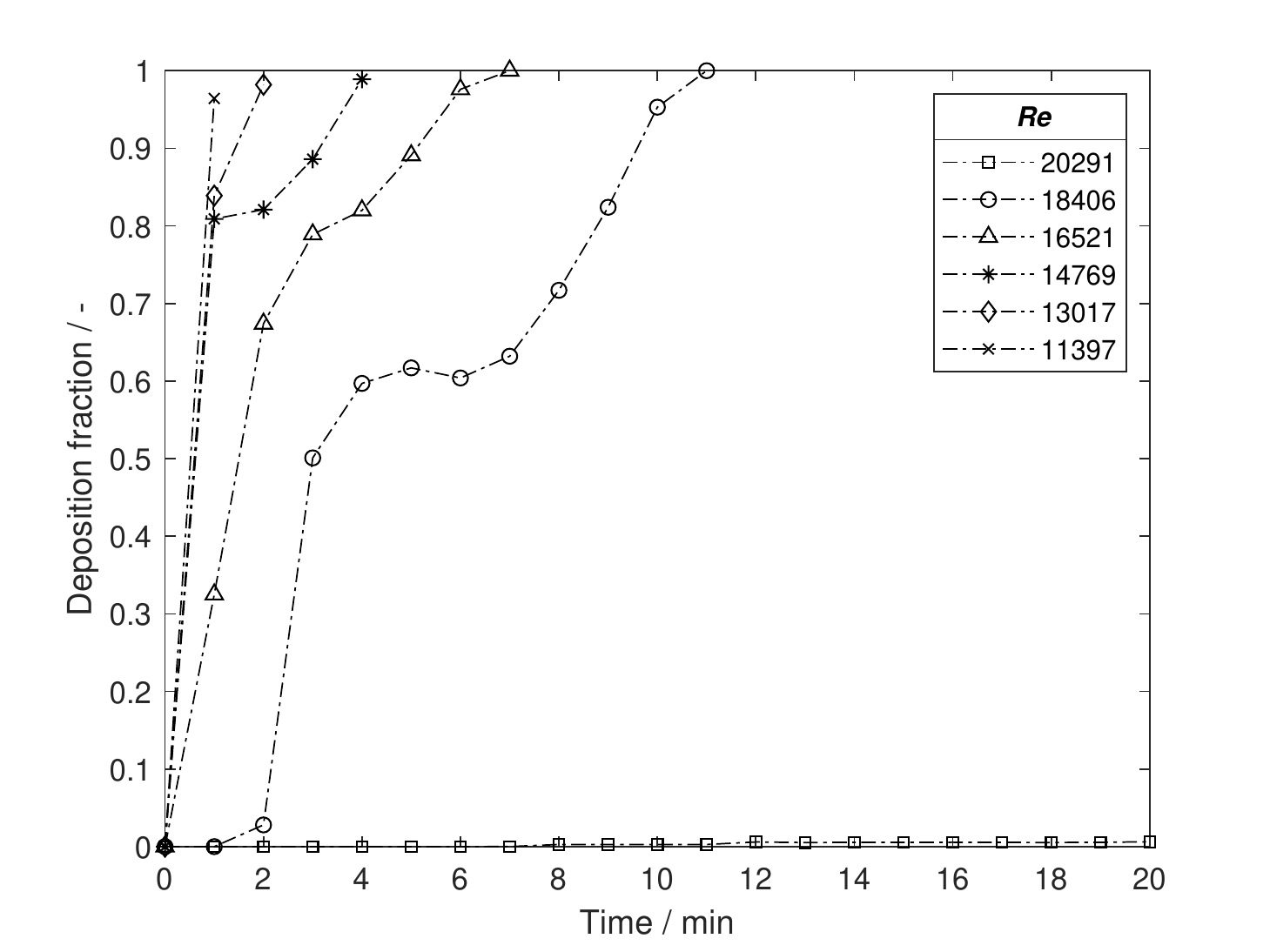}\label{Poly2}}\\
\subfigure[]{\includegraphics[width=.47\textwidth]{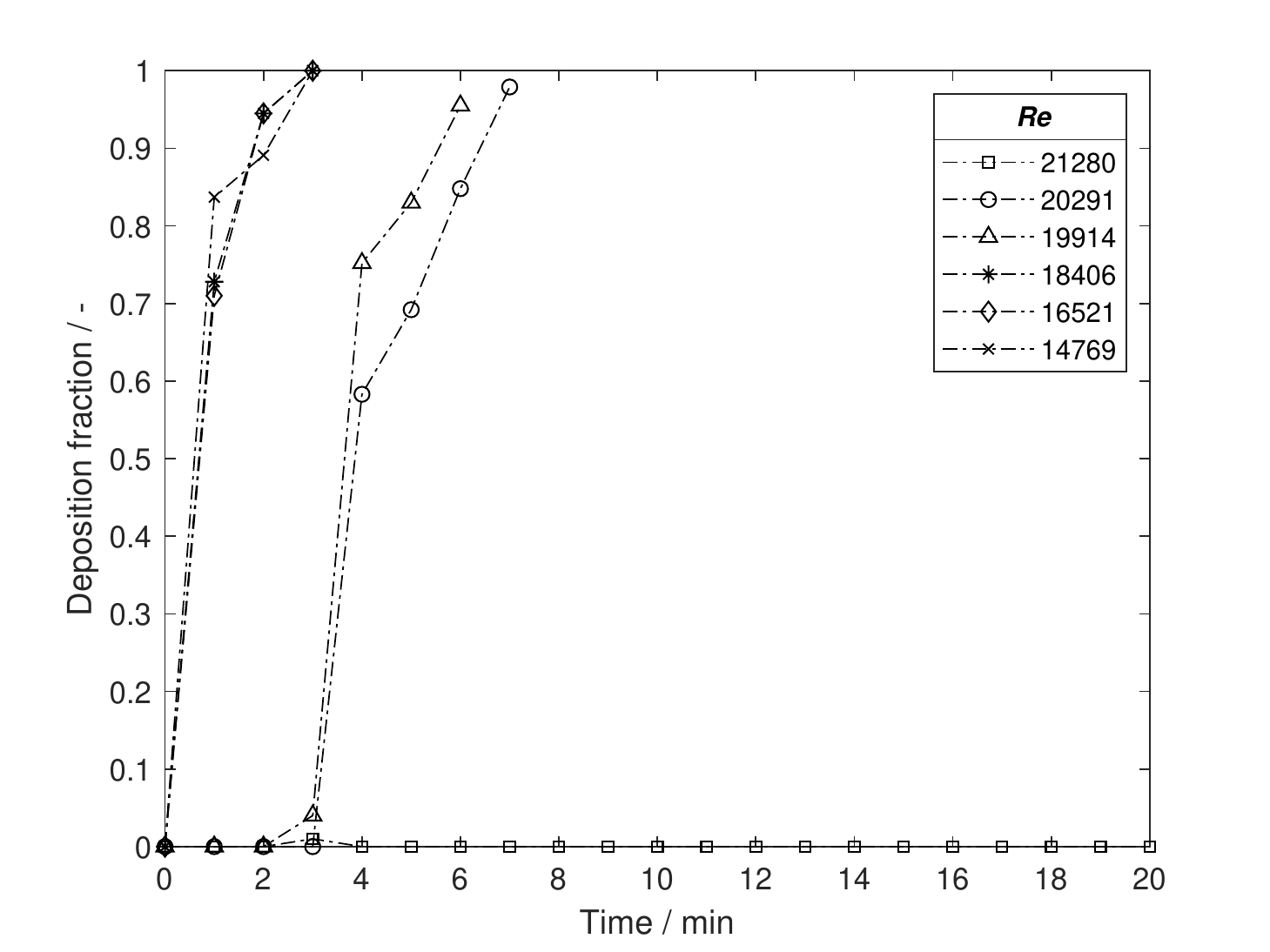}\label{Poly3}}\quad
\subfigure[]{\includegraphics[width=.47\textwidth]{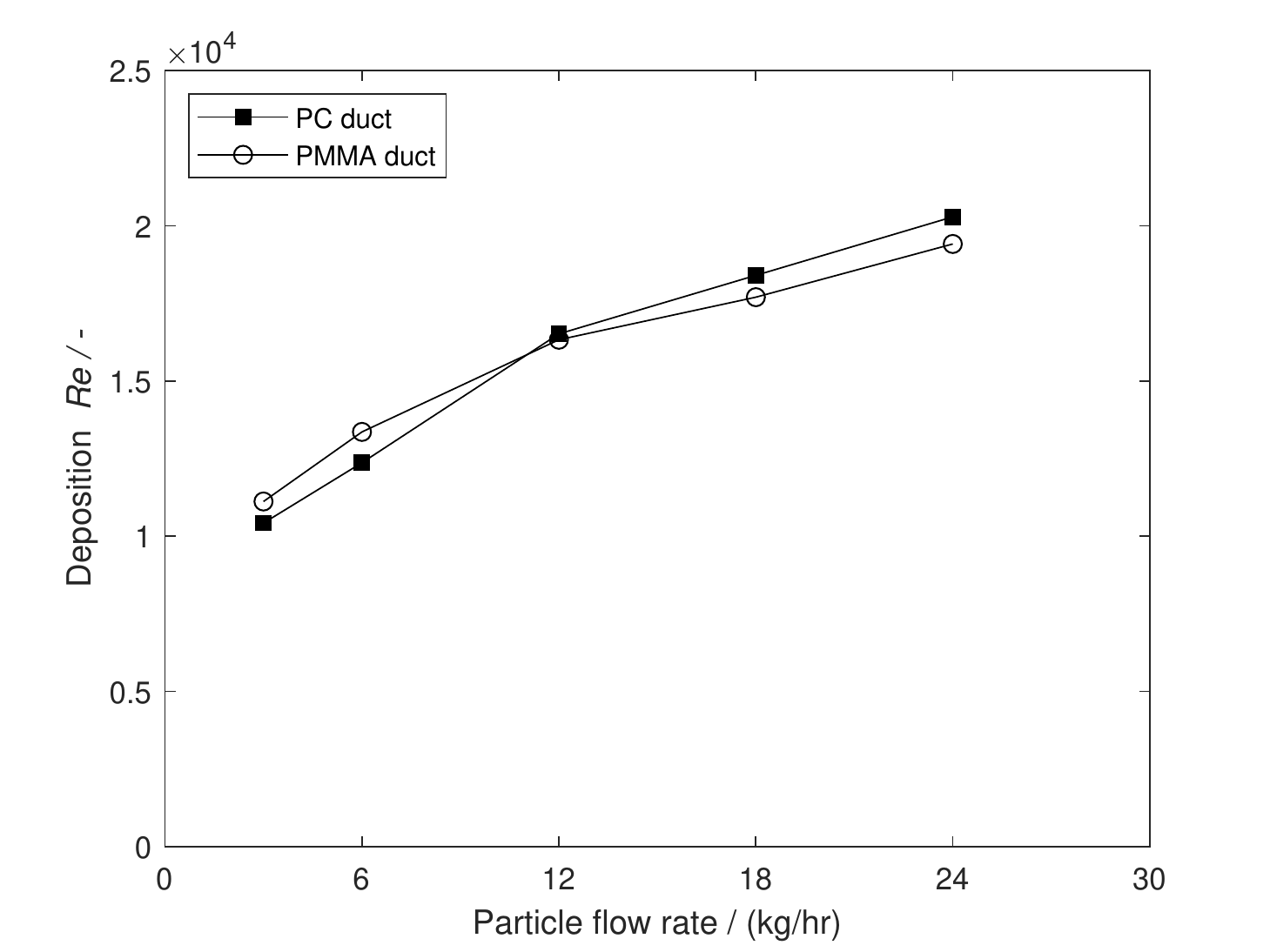}\label{Deposition Velocity Flowrate}}
\caption{Particle deposition for different flow Reynolds numbers as a function of time for a particle mass flow rate of (a) 3~kg/hr, (b) 12~kg/hr, and (c) 18~kg/hr. 
In (d) the deposition velocity based on a conveying duration of 1 min is summarized.}
\label{depositionfxtime}
\end{figure*}

For polydisperse particles, we measured the instantaneous particle deposition fraction for different flow $Re$ numbers keeping the particle flow rate constant during each experiment while varying the flow rate in different experiments between 3 kg/hr and 24 kg/hr.
Thus, a dilute conveying phase is present.
Each test was performed for 20 min using the PMMA duct and RH ranging between 30\% and 36\%.

The results presented in Fig.\,\ref{depositionfxtime} show that by reducing \(Re\) more particles settle at the pipe bottom and the deposition process occurs faster.
Thus, a layered bed is created which at a certain time will block the cross-sectional area of the duct and prevent the particles to flow.
As mentioned above, close to the deposition velocity the flow is highly unstable.
This process is optically clearly observable at higher particle flow rates where the particles are no longer homogeneously distributed, but rather resemble particle clusters sliding at the bottom of the pipe.

Finally, the resulting deposition velocity for all experiments in both a PC and PMMA duct are summarized in Fig.~\ref{Deposition Velocity Flowrate}.
As expected, the deposition velocity increases with increasing particle flow rate.
In essence, the more particles are present in the conveying air, the higher the required air velocity to keep the particle in suspension.

Moreover, the usage of transparent ducts enables us to ascertain that particles that do not have enough inertia to propagate to the discharge point halt once they hit the bottom wall of the duct.
The following particles then hit these initially settling particles and join together resulting in a lengthwise monolayer deposit bed.
Thus, the initial settling particle acts seed-like as an initial point where the deposit starts to grow.
Meanwhile, the particles above the bed are still transported in the dilute phase.
As the deposit bed increases its length towards the feeding point, the following particles no longer hit the bottom of the pipe, but the deposit layer.
These particle collisions decrease the particle velocity.
Thus, they are no longer in suspension, rather rolling above the deposit bed and slowly adding up the bed height into a multilayer deposit until it is sufficient to completely block the duct.

As can be seen in Fig.~\ref{Poly1}, the deposition velocity of polydisperse particle fed at a rate of 3 kg/hr is 5.0~m/s ($Re$~=~14769).
However, deposition is also observed at 5.6~m/s ($Re$~=~16521) related to a different mechanism.
Before 5 min deposition forms and afterward the particles start to drop out of suspension and generate a tail pattern of monolayer deposit at the center part of the bottom duct.
The following particles hit these deposited particles; however, they do not join together as with the lower $Re$.
We could observe that the airborne particles swept away the particles in the deposit and took instead their place.
Thus, an equilibrium of deposited and resuspended particles is created and, in sum, the amount of deposit is constant.
This type of equilibrium behavior was reported by other authors before \citep{Adhiwidjaja2000,Matsusaka1998}.

\subsection{Deposition shape and location}

\begin{figure*}[b]
\centering
\subfigure[Polydisperse, $Re$~=~12705]{\includegraphics[trim=0cm 0cm 0cm 0cm,clip=true,width=.4\textwidth]{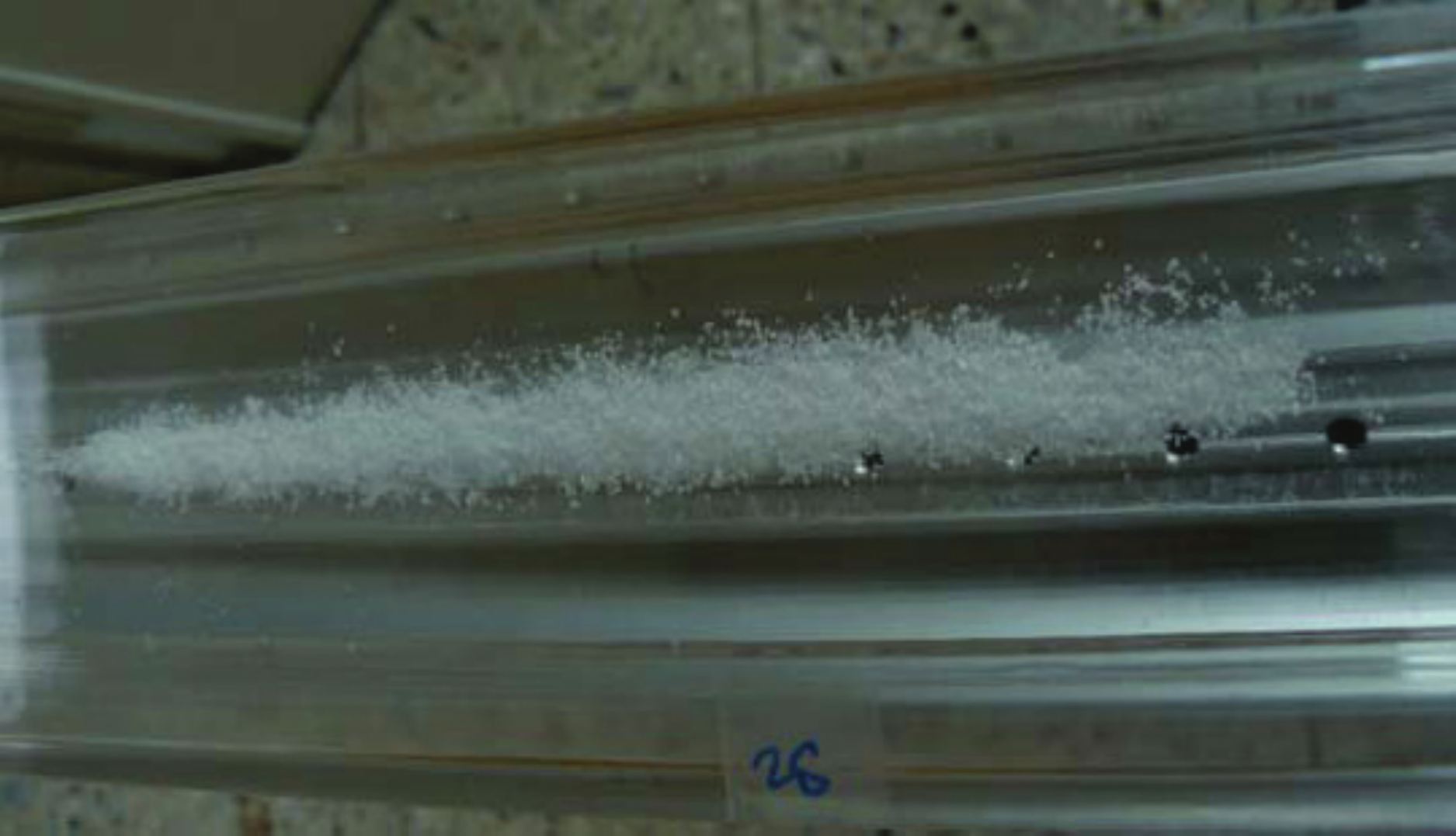}\label{Polypicthigh}}\qquad
\subfigure[Polydisperse, $Re$~=~11437]{\includegraphics[trim=0cm 0cm 4cm 0cm,clip=true,width=.4\textwidth]{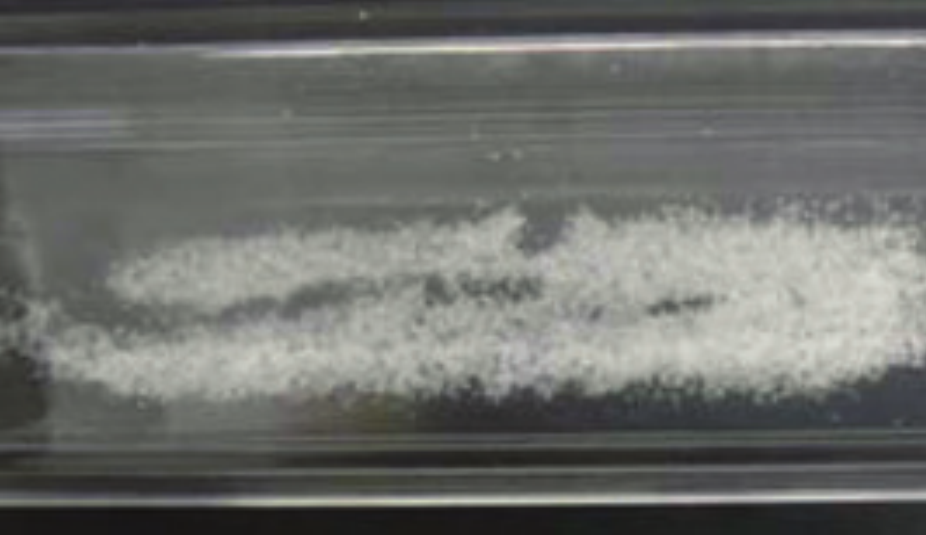}\label{Polypictlow}}\\
\medskip
\caption{Deposition patterns in the PMMA duct.
The airflow in (a) is slightly above the deposition velocity and in (b) slightly below.}
\label{PictRe}
\end{figure*}

In Fig.~\ref{PictRe} typical deposit patterns for two different Reynolds numbers are presented.
Therein, the left picture relate to a flow slightly above the respective deposition velocity and the right one slightly below.
As can be seen, the particles tend to deposit at the center of the duct.

This is likely related to fluid mechanical phenomena:
it is known that the concentration  of particles in a wall-bounded flow is inhomogeneous due to the turbophoretic drift under the influence of near-wall coherent turbulent structures.
Also, the squared-shaped duct under consideration has two inhomogeneous directions (wall-normal and spanwise) which induces a secondary flow of Prandtl's second kind \citep{Gessner1973,Yao2010}.
As a result of this mechanism, four pairs of counter-rotating vortices are generated in the square duct \citep{Wu2018}.
This secondary flow affects particle transport and dispersion,  and consequently, deposition location.
It enhances lateral mixing which is found stronger at the corner of the duct and supposedly responsible for the particle spanwise motion and transporting particles from the duct corner to the duct center \citep{Sharma2006}.
However, that motion considerably weakens in the duct center.
Here, streamwise turbulence dominates the flow where turbophoresis occurs and brings the particles again to the region having low turbulence intensity, which is in the duct corner.
These visual evidences confirm the results of several numerical studies \citep{Noo16,Winkler2006,Sharma2006,Lin2017,zhang2015}.

Further, based on the Reynolds number, the shape of the deposit appears differently.
Analyzing the deposition shape of the lower flow $Re$ depicted in Fig.\,~\ref{Polypictlow} it can be seen that the streaky pattern looks more dispersed compared to concentrated ones of the higher $Re$ case in Figs.\,~\ref{Polypicthigh}.
When the shear rate is decreased, the near-wall vortices are no longer able to eject and entrain particles.
As a result, they will only be pushed aside by the sweep motion.
On the contrary, when the shear rate is increased, vortices have sufficient energy to entrain particles causing the thin streaky deposition pattern \citep{Kaftori1995}.

\begin{figure}[tb]
        \centering
        \includegraphics[width=.47\textwidth]{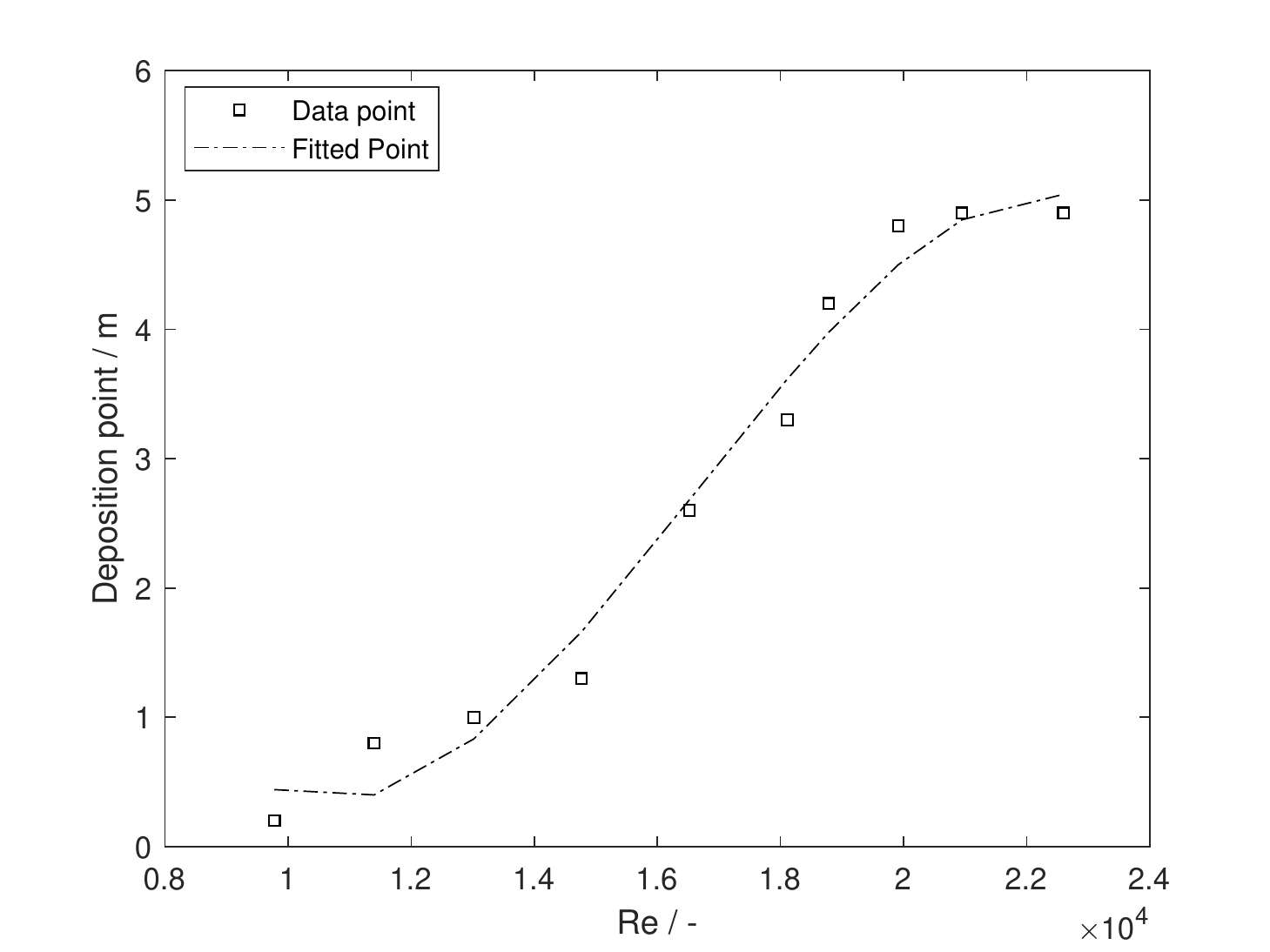}
        \caption{Influence of the air velocity on the deposition point of the polydisperse particles.}
        \label{DepositionPoint}
\end{figure}

Moreover, we investigated the influence of the air velocity on the deposition point.
The deposition point is the point at which particles start to deposit and create a layer measured as the distance from the feeding point.
For that analysis, a PC duct with a total length of 5.4 m was used.
As shown in Fig.\,\ref{DepositionPoint}, the increasing air velocity is able to convey particles a farther distance until a constant point is reached.
In other words, a higher air velocity will no longer influence the transport distance.
This point is also called the boundary saltation.

\section{Conclusion}

The presence of particle deposits during pneumatic conveying is a potential hazard and ignition source for dust explosions.
The objective of this work is to design and build a new test-rig to investigate particle deposition phenomena in squared ducts.
Further, we analyzed the influence of the air conveying velocity, particle mass flow rate, relative air humidity and temperature, and duct material.
It was found that the higher the particles' flow rate, the higher the required air velocity to transport the powder to the outlet.
High RH caused small particles to deposit at higher air conveying velocity and increases the particle deposition fraction.
These findings enhance our understanding of the detailed mechanisms underlying the formation of deposits.

\section*{Acknowledgements}

The authors gratefully acknowledge the financial support from Max Buchner Research Foundation.

\bibliography{../../../publications/publications.bib}

\end{document}